# A macroscopic delayed-choice quantum eraser using a commercial laser

Byoung S. Ham

School of Electrical Engineering and Computer Science, Gwangju Institute of Science and Technology,
123 Chumdangwagi-ro, Buk-gu, Gwangju 61005, S. Korea
(June 11, 2024; bham@gist.ac.kr)

**Abstract**

The heart of quantum mechanics is quantum superposition between orthogonal bases of a single particle. In the particle nature of quantum mechanics, quantum superposition is represented by probability amplitudes between mutually exclusive natures such as orthogonal polarization bases. The delayed-choice quantum eraser is for the post-determination of the photon's nature, raising the cause-effect relation issue. Over the last several decades, quantum erasers have been intensively studied using nearly all kinds of photons. Here, the macroscopic delayed-choice quantum eraser is experimentally demonstrated using a continuous wave laser and discussed for quantum superposition in a macroscopic regime. For this, a noninterfering Mach-Zehnder interferometer composed of two polarizing beam splitters is chosen to manipulate polarization bases of lights and to measure them in a delayed-choice manner via polarization-basis projection.

**Introduction**

Quantum superposition between random bases of a physical entity such as a single photon or atom in an interferometric system is the heart of quantum mechanics as mentioned by Dirac [1] and Feynman [2]. One of the fundamental roles of measurements is in controlling the bases, resulting in the mysterious quantum features of delayed choice [3] and a quantum eraser [4-8]. As a result, the cause-effect relation has been an important issue due to the retrospective choice between the wave and particle natures. The delayed-choice phenomenon is for the complementarity between which way information and the fringe visibility of a physical entity in a noninterfering interferometer. According to the interpretation of quantum mechanics [1-3], the which-way information of a physical entity relates to the particle nature of distinguishability, prohibiting interference fringes. Complete randomness on which way information results in perfect fringe visibility of the wave nature for indistinguishability. These two quantum properties are mutually exclusive resulting in the wave-particle duality [1]. Thus, the quantum mystery of the delayed choice is in the retrospective selection between the particle nature with perfect which-way information and the wave nature with complete randomness to the which way. In terms of the post-action of the delayed measurement, the quantum eraser raises the causality issue as proposed by Scully and Drühl in 1982 [4,5] and demonstrated by Kim et al. in 2000 [8].

In this paper, a coherence approach is conducted for a macroscopic quantum eraser using a continuous-wave (cw) laser in a noninterfering Mach-Zehnder interferometer (NMZI). The NMZI predetermines the photon's nature as a particle, prohibiting interference fringes. For the delayed choice, however, the polarization bases of NMZI output fields are coherently controlled using a polarizer via polarization-basis projection measurements [9]. Due to the equality between quantum and coherence approaches for the first-order intensity correlation [10], there is no fundamental difference between a single photon and cw light for the MZI fringes [11-14]. This is originated in the thumb rule of quantum mechanics that a photon never interferes with others [1]. In other words, the cw NMZI fringe is nothing but incoherent collections of the single photon's self-interference [12]. The first proof of this fact was conducted using entangled photons in 1986 [14]. Over the last few decades, delayed-choice quantum-eraser experiments have been intensively conducted in a microscopic regime using thermal light [15], coherent photons [16-18], antibunched photons [19,20], entangled photons [7,8,21], and even spin particles [22]. Recently, the cw quantum eraser has been applied for superresolution [23,24] in quantum sensing [25,26] using the same projection measurement [9]. Such a projection measurement has been a major tool in nonlocal quantum correlation for entangled photon-pair generation [27] and the Bell inequality violation [28,29]. Thus, the correct understanding of the quantum eraser is essential to various quantum technologies [1-4].

**Experimental model of the macroscopic quantum eraser**

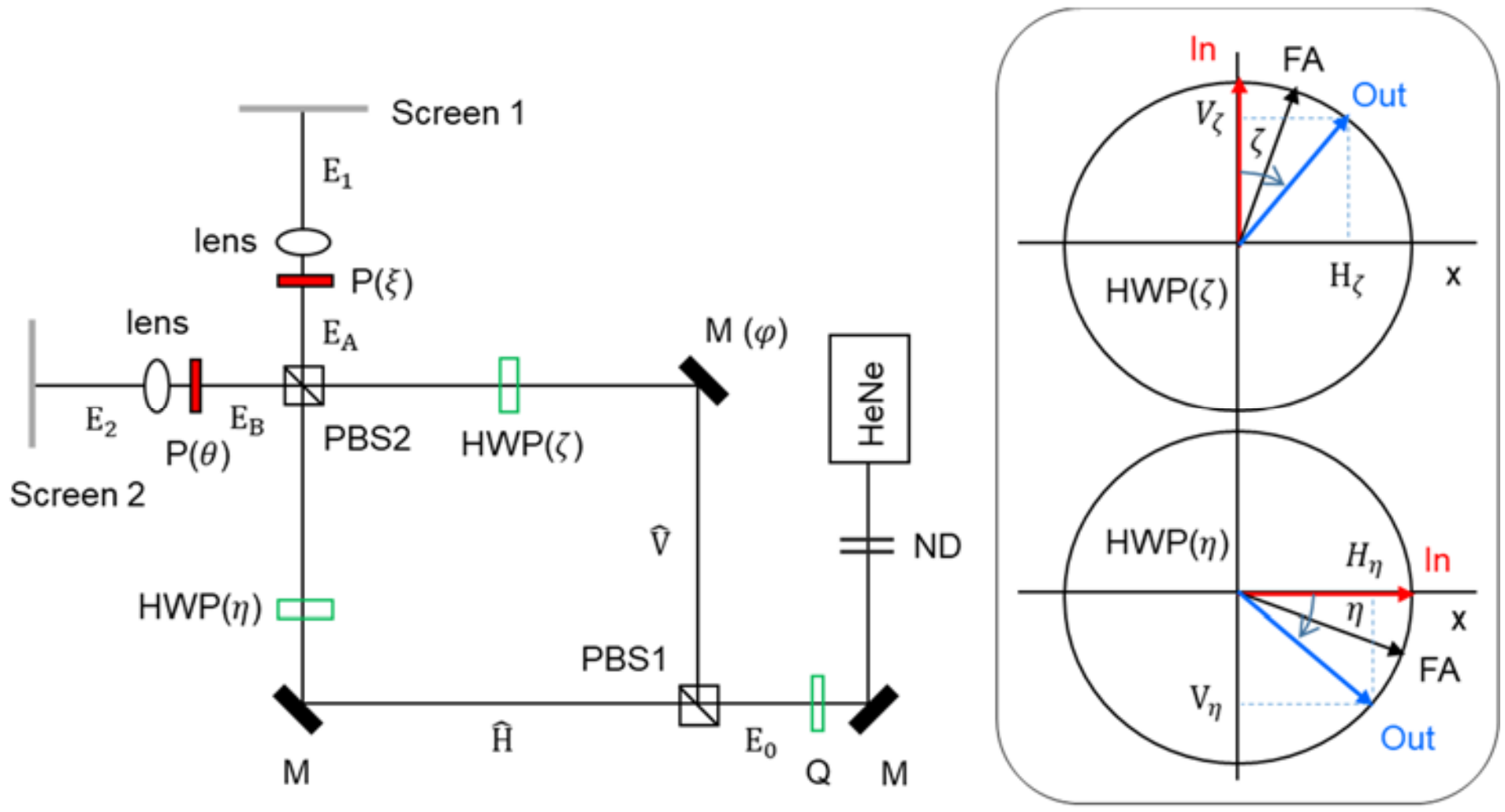

**Fig. 1.** Schematic of Wheeler's thought experiments of quantum eraser using coherent light. ND: neutral density filter, M: mirror, Q: quarter-wave plate, PBS: polarizing beam splitter, HWP: half-wave plate, P: polarizer. Inset: HWP control convention. FA: fast axis of HWP.

Figure 1 shows the schematic of the present macroscopic quantum eraser in NMZI composed of PBSs. For the random polarization bases of light, a 45-degree rotated quarter-wave plate (Q) is placed just before NMZI, where a relative phase between perpendicularly polarized lights has an effect of phase shift [23,24]. By PBS1, the photon's nature is preset as a particle with perfect which-path information in the NMZI path. A 22.5-degree rotated half-wave plate (HWP) is placed in each NMZI path to randomly generate vertically and horizontally polarized photons (see Inset for the rotation convention). By the PBS2, however, no interference fringe is generated in the output fields due to the Fresnel-Arago law [11], where the random bases of the diagonally and antidiagonally polarized photons are selectively chosen by PBS2 for only either vertical or horizontal components. Finally, the polarizer P followed by PBS2 acts for the test of the quantum eraser. Here, the role of Ps is the polarization-basis projection of the distinguishable photons onto the rotation angle of $\xi$ and $\theta$. The input light in Fig. 1 is from a TEM00 single-mode cw HeNe laser (Thorlabs HRS015B) at the wavelength $\lambda = 633$ nm, whose linewidth and polarization are 1 MHz and vertical, respectively. The input power $I_0$ $(E_0E_0^*)$ of the HeNe laser is set at ~100 μW before entering NMZI.

**Theory**

The output fields of NMZI in Fig. 1 is represented using the BS matrix representation [30]:

$$\boldsymbol{E}_A = \frac{E_0}{2}\left(H_\eta - V_\zeta e^{i\varphi}\right), \tag{1}$$

$$\boldsymbol{E}_B = \frac{iE_0}{2}\left(V_\eta + H_\zeta e^{i\varphi}\right), \tag{2}$$

where $V_j$ and $H_j$ indicate unit vectors of the HWP-resulting random polarization components, as indicated in the Inset of Fig. 1. By the rotation angles of HWPs, $H_\zeta = \hat{V}sin\zeta$, $V_\zeta = \hat{V}cos\zeta$, $H_\eta = \hat{H}cos\eta$, and $V_\eta = \hat{H}sin\eta$ are resulted by PBS2. Convention of the rotation angle $\zeta$ and $\eta$ is shown in the Inset of Fig. 1. The term $e^{i\varphi}$ is the NMZI path-length difference-caused phase by controlling the V component only, where the Q-induced additional phase is included. From Eqs. (1) and (2), no interference fringe results in the output fields $I_A$ and $I_B$, testifying no interaction between orthogonal polarization bases [13]. This coherence analysis of NMZI in Eqs. (1) and (2), representing the particle nature of quantum mechanics, can also be interpreted using quantum operators in the same way [31].

If PBS2 is replaced by a 50/50 nonpolarizing beam splitter (BS), NMZI results in fringes in both $I_A$ and $I_B$ with perfect fringe visibility, as numerically calculated in the left column of Fig. 2 [32]: Details are given in the Discussion section. This is the original concept of the delayed-choice thought experiment proposed by Wheeler in 1978, where HWPs play the role of the delayed choice [3]. In this case, the replaced BS acts as a two-input, two-output interferometer, whose input fields are perfectly coherent with random polarization bases generated by HWPs. By the way, without HWPs in Fig. 1, the output intensities become $I_A = I_0$ and $I_B = 0$ due to $V_h = H_v = 0$ by PBS2 (see the second column in Fig. 3). These are experimentally confirmed (not shown).

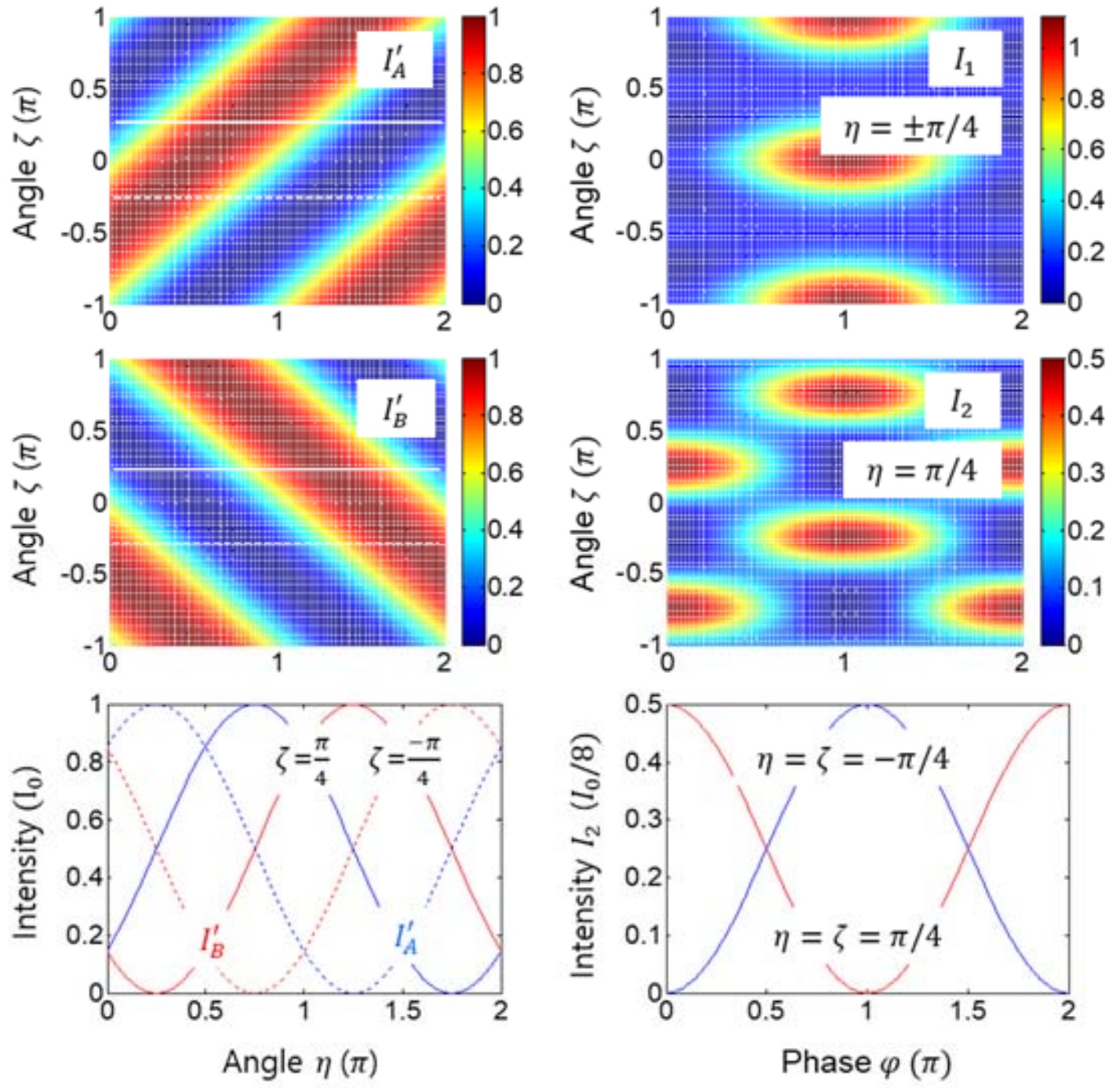

**Fig. 2.** Numerical simulations for Fig. 1. (left column) PBS2 is replaced by BS with $\varphi = 0$. Solid (dotted) curve: $\zeta = \frac{\pi}{4}$ ($\zeta = -\frac{\pi}{4}$). (right column) For Eqs. (5) and (6). $\zeta = \theta = \pi/4$.

By the polarizers Ps in Fig. 1, the polarization-basis projection of the output fields $E_A$ and $E_B$ onto the polarizer is carried out. The resulting amplitudes of the output fields are obtained from Eqs. (1) and (2):

$$E_1 = \frac{E_0}{2}\left(H_\eta cos\xi - V_\zeta sin\xi e^{i\varphi}\right), \quad (3)$$

$$E_2 = \frac{iE_0}{2}\left(V_\eta sin\theta + H_\zeta cos\theta e^{i\varphi}\right), \quad (4)$$

where $\xi$ and $\theta$ are rotation angles of Ps from the horizontal axis. Due to the P-caused polarization projection of the orthogonal bases in each output field onto the rotated Ps, the corresponding intensities of Eqs. (3) and (4) are as follows for $\zeta = \eta = \frac{\pi}{4}$ (D) of HWPs:

$$I_1 = \frac{I_0}{8}(1 - sin2\xi cos\varphi), \quad (5)$$

$$I_2 = \frac{I_0}{8}(1 + sin2\theta cos\varphi), \quad (6)$$

where $H_\zeta = H_\eta = V_\eta = V_\zeta = 1/\sqrt{2}$. In Eqs. (5) and (6), the notation with H and V is for a unit vector to clarify the origin of coherence retrieved by Ps. Thus, the HV product-basis term becomes effective now, resulting in interference fringes, as shown in the right column of Fig. 2 (see also the third and fourth columns in Fig. 3). This is the quintessence of the macroscopic quantum eraser for the projection measurements.

Similarly, the following relations can be derived from Eqs. (3) and (4) for fixed Ps at $\xi = \theta = \frac{\pi}{4}$:

$$I_1 = \frac{I_0}{8}\left[(cos^2\eta + cos^2\zeta) - 2cos\eta cos\zeta cos\varphi\right], \quad (7)$$

$$I_2 = \frac{I_0}{8}[(sin^2\eta + sin^2\zeta) + 2sin\eta sin\zeta cos\varphi]. \quad (8)$$

For the orthogonal bases of light by the same diagonally $\left(D;\ \frac{\pi}{4}\right)$ or antidiagonally $\left(A;\ -\frac{\pi}{4}\right)$ rotated HWPs (η; ζ), Eqs. (7) and (8) are rewritten as $I_{1,2} = \frac{I_0}{8}(1 \mp cos\varphi)$. For opposite HWPs ($\zeta$: $D$; $\eta$: $A$), however, the fringe in Eq. (8) is reversed (see the bottom-right panel in Fig. 2 and the fourth column in Fig. 3): $I_2 = \frac{I_0}{8}(1 - cos\varphi)$. $I_1$ is independent of the D or A of HWPs (see top-right panel in Fig. 2). This is due to $H_\zeta = V_\eta = 0$ in Path A for $I_1$ by PBS2. Thus, the P's selective measurement for the polarization basis of the NMZI output fields results in the macroscopic quantum eraser. The fringe swapping in $I_2$ happens only for opposite basis selections of HWPs.

**Experimental results**

Figure 3 shows unprecedented experimental demonstrations of the macroscopic delayed-choice quantum eraser in Fig. 1. The images in Fig. 3 are captured from both screens S1 and S2 in Fig. 1. For the real-time comparison, $I_1$ was intercepted by a mirror and sent to Screen S2. For visualization purposes, a spatial misalignment of lights on PBS2 is intentionally applied for the fringe formation. The first column from the left shows the NMZI output field's intensities $I_A$ and $I_B$ without Ps, where no interference fringe results in regardless of HWPs (see Eqs. (1) and (2)). As analyzed in the Theory section, thus, the photon characteristics of NMZI confirm the particle nature. The HWP-dependent intensities are due to the polarization projection onto PBS2. Without HWPs, $I_B = 0$ results in because the vertical (horizontal) component of the H- (V-) polarized photon is zero, $V_x = H_y = 0$, as shown in the second column. If the PBS2 is replaced by BS, the photon characteristics of the wave nature are retrieved, resulting in fringes (not shown), as analyzed in the left column of Fig. 2.

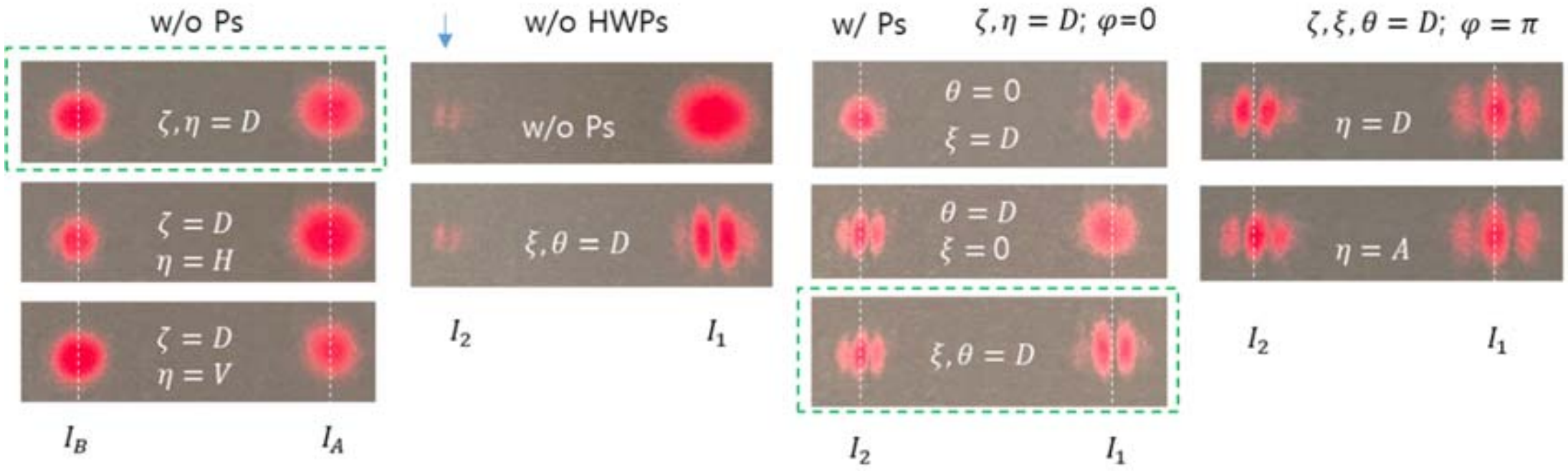

**Fig. 3.** Experimental demonstrations of delayed-choice for the quantum eraser. $I_j$ is the intensity of $E_j$. A (D) is for the anti-diagonal (diagonal) in P's rotation from the horizontal axis into a counterclockwise direction.

The second column in Fig. 3 shows the output field's images without HWPs. Thus, the photon inside the NMZI is preset for the particle nature by PBS1 with perfect which-way information. The top panel is without polarizers, resulting in $I_2 = 0$ due to $V_x = H_y = 0$, as mentioned above. Thus, all light travels to output Path A. The dim images of $I_2$ are caused by imperfect PBSs, where the unwanted leakage of polarized lights (~1 %) results in the dim interference fringes in $I_2$. The fringe variation was also confirmed with φ variation (not shown). As expected for the indistinguishable photon characteristics, no fringe is observed in $I_1$ without P, resulting in no quantum eraser. The bottom panel is with Ps, whose polarization axis is aligned along the diagonal direction (D), resulting in the retrieval of interference fringes of the macroscopic quantum eraser. This is the first observation of the delayed-choice quantum eraser in a macroscopic regime without HWPs for Fig. 1. Thus, the physics of the quantum eraser is confirmed for the photon's polarization-basis control by P as a measurement choice.

The third column is for the experimental proof of the macroscopic delayed-choice quantum eraser with HWPs for Fig. 1. Using polarizers is a typical measurement technique in the Bell inequality violation (see Discussion) [7,28,29]. The top panel is with a polarizer (P) only applied to path A for S1, resulting in the retrieval of interference fringes in $I_1$ only. The middle panel is for the swapped case of the top panel by relocating the polarizer from path A to B, resulting in switching fringe to $I_2$. The bottom panel is for polarizers put in both paths, resulting in interference fringes in both $I_1$ and $I_2$. Thus, the role of the polarizer was demonstrated for the delayed-choice quantum eraser, where the particle nature set by PBS2 is converted to the wave nature in a time-reversed manner. Compared with the top panel of the first column, nothing changes in the bottom panel of the third column except the use of polarizers in the output ports (see the green dashed boxes).

The last column is for the control of HWPs and $\varphi$. The top panel is for the same HWPs rotated by $22.5°$ $(\eta = \zeta = 45°)$ as a reference, where $\varphi = \pi$ is set. Compared with the bottom panel of the third column, $I_1$ and $I_2$ are swapped according to Eqs. (5) and (6) for $\varphi$. The bottom panel is for the anti-diagonal direction (A) of one HWP $(\eta = -45°;\ \zeta = +45°)$ applied to $\eta$ only, resulting in fringe swapping only in $I_2$. As coherently derived in Eqs. (7) and (8), the fringe control is due to $\sin\eta\sin\zeta$ for $I_2$, where $I_1$ has no such components by PBS2 (see Eqs. (1)-(4)). When the $\zeta$-HWP is controlled between D and A for a fixed $\eta$, the fringe swapping was confirmed in $I_2$ only (not shown). Therefore, HWPs correlate with output fringes via Ps. This explains the polarization basis control by Ps for measurement choices. In other words, the preset orthogonal polarization bases by PBS1 (PBS2) can be retrospectively converted to be random by HWP (P).

The output fringes are also $\varphi$-dependent in Fig. 3. Depending on $\varphi \in \{0, \pi\}$, both output fringes show swapping patterns (see the bottom panel of the third column and the top panel of the last column). If the path-length difference of the NMZI is beyond the coherence length of the laser, no delayed-choice quantum eraser is observed [32]. This means that even the concept of the particle nature in quantum mechanics must imply coherence of the wave nature for the quantum feature. Thus, the particle nature of quantum mechanics must be differentiated from the incoherent classical particles. Without coherence, no quantum eraser exists.

## Discussion

*Interpretation of the delayed choice and ad-hoc quantum superposition*

If BS replaces PBS2 in Fig. 1, the NMZI output fields $I_A$ and $I_B$ show perfect interference fringes, as numerically calculated in the left column of Fig. 2 [32]. This is for the original delayed-choice thought experiment proposed by Wheeler in 1978 [3]. In this modified scheme, the particle nature set by PBS1 for the perfect which-way information is retrospectively switched to the wave nature by the delayed choice of HWP. The heart of Wheeler's delayed-choice thought experiment is regard to quantum superposition and a measurement choice. As demonstrated in many delayed-choice experiments with nearly all kinds of photons [15-21,32], the observed macroscopic quantum feature in Fig. 3 also shows the same quantum mystery how the particle nature predetermined by PBS1 can be converted into the wave nature by HWP in a time delayed manner. Reminding that a cw interference fringe is an incoherent accumulation of each single photon's self-interference [1,2,12,14], the origin of the macroscopic delayed-choice quantum eraser observed in Fig. 3 is in the single photon's case [32].

Like the polarization-basis projection onto the polarizer observe in Fig. 3, the HWP-provided random polarization bases are projected onto the replaced BS in Fig. 4. According to coherence optics [33], HWP results in a $\pi$ phase shift to the vertical component with respect to the horizontal one due to the birefringence effect, resulting in the $\zeta$ ($\eta$) rotated linearly polarized light 'Out' (see also the Inset of Fig. 1). Thus, the *ad-hoc* orthogonal polarization components of the light ‘Out’ from each HWP are superposed on the BS (see the red and blue arrows). In coherence optics, such decomposition of the rotated field ‘Out’ into the vertical and horizontal ones has been commonly accepted without a doubt [33]. Like the single-photon interference in MZI [12,14], however, the same dilemma is confronted due to the unbreakable minimum energy of a single photon. Thus, the quantum superposition of the orthogonal bases for ‘Out’ is also mysterious in a microscopic regime, where the reduced projection lengths stand for the probability amplitudes of finding.

In Fig. 4, $H_\zeta$ is parallel to $H_\eta$. Similarly, $V_\eta$ is antiparallel to $V_\zeta$. For $\zeta = \eta = \pi/4$, their magnitudes are the same as $I_{out}/2$. Thus, the same-colored projected photons or lights in each pair of HWPs cannot be distinguishable on BS due to no difference phase in each color set. As a result, the interference fringe is formed in the output ports of BS from this amplitude superposition between the *ad-hoc* orthogonal bases (see the left column of Fig. 2):

$$E_A^{'} = \frac{1}{\sqrt{2}}\left[\left(-V_\zeta e^{i\varphi} - V_\eta\right)\hat{V} + \left(-H_\zeta e^{i\varphi} + H_\eta\right)\hat{H}\right],$$
$$= -\frac{E_0}{2}\left[(cos\zeta e^{i\varphi} + sin\eta)\hat{V} + (sin\zeta e^{i\varphi} - cos\eta)\hat{H}\right], \tag{11}$$
$$E_B^{'} = \frac{i}{\sqrt{2}}\left[\left(\hat{H}_\zeta e^{i\varphi} + V_\eta\right)\hat{V} + \left(H_\zeta e^{i\varphi} + H_\eta\right)\hat{H}\right],$$
$$= \frac{iE_0}{2}\left[(cos\zeta e^{i\varphi} + sin\eta)\hat{V} + (sin\zeta e^{i\varphi} + cos\eta)\hat{H}\right]. \tag{12}$$

From Eqs. (11) and (12), corresponding interference fringes result, demonstrating the delayed choice thought experiments in a macroscopic regime (see the numerical calculations in the left column of Fig. 2):

$$I_A^{'} = \frac{I_0}{4}\left[(cos\zeta e^{i\varphi} + sin\eta)(cos\zeta e^{-i\varphi}sin\eta) + (sin\zeta e^{i\varphi} - cos\eta)(sin\zeta e^{-i\varphi} - cos\eta)\right],$$
$$= \frac{I_0}{4}\left[(cos^2\zeta + sin^2\eta + 2sin\eta cos\zeta cos\varphi) + (cos^2\eta + sin^2\zeta - 2sin\zeta cos\eta cos\varphi)\right],$$
$$= \frac{I_0}{2}\left[1 + \sin(\eta - \zeta)cos\varphi\right]. \tag{13}$$
$$I_B^{'} = \frac{I_0}{4}\left[(cos\zeta e^{i\varphi} + sin\eta)(cos\zeta e^{-i\varphi} + sin\eta) + (sin\zeta e^{i\varphi} + cos\eta)(sin\zeta e^{-i\varphi} + cos\eta)\right],$$
$$= \frac{I_0}{4}\left[(cos^2\zeta + sin^2\eta + 2sin\eta cos\zeta cos\varphi) + (cos^2\eta + sin^2\zeta + 2sin\zeta cos\eta cos\varphi)\right],$$
$$= \frac{I_0}{2}\left[1 + \sin(\eta + \zeta)cos\varphi\right]. \tag{14}$$

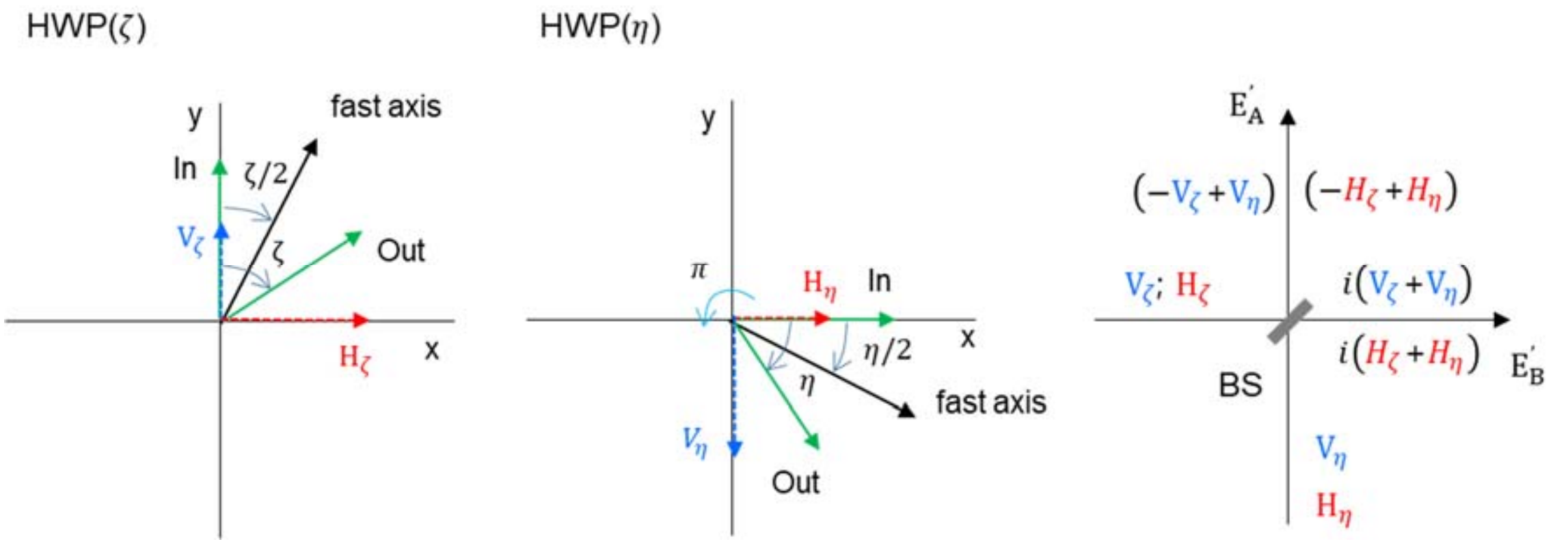

**Fig. 4.** Schematic of polarization-basis control by HWPs for ad-hoc quantum superposition.

As Wheeler mentioned [3], the delayed choice in Fig. 4 for Eqs. (13) and (14) should invoke the same causality issue. As demonstrated with coherent single photons [32], however, the delayed choice quantum eraser observed in the third column of Fig. 3 for a macroscopic regime is also critical to the coherence length of the cw light. Concerning the causality issue of the delayed choice between PBS1 and HWP in Figs. 2 and 4 or PBS1 and Ps in Figs. 1 and 3, the action distance beyond the light cone has nothing to do with coherence. This is because coherence is for a relative concept between two objects, whereas the light cone is an absolute one in each object. Thus, the causality issue in the present delayed-choice quantum eraser is not appropriate due to the coherence approach. The delayed choice by HWP or P observed in Fig. 3 results from the selective measurements out of all possible combinations via polarization-basis projection. For this, the *ad-hoc* quantum superposition plays an important role, as analyzed in Eqs. (13) and (14). In other words, any polarization component of a photon or light implies infinite number of sets of orthogonal bases.

*Nonlocal correlation via selective measurement in quantum erasers*

Figure 5 is for the second-order intensity correlation $R_{12}(0)$ between two output fields in Fig. 1, where the 'Screens' are replaced by photodetectors. For the coherently excited nonlocal correlation, only $H_\eta H_\zeta$ and $V_\zeta V_\eta$ product-basis terms are selectively allowed for $R_{12}(0)$ by any reason to mimic entangled photon pairs [21,26,27,29]:

$$R_{12}(0) = \frac{I_0^2}{16}\left(H_\eta cos\xi - V_\zeta sin\xi e^{i\varphi}\right)\left(V_\eta sin\theta + H_\zeta cos\theta e^{i\varphi}\right)(cc),$$

$$= \frac{I_0^2}{16} e^{i\varphi}\left(H_\eta H_\zeta cos\theta cos\xi - V_\zeta V_\eta sin\theta sin\xi\right)(cc),$$

$$= \frac{I_0^2}{16}\left(H_\eta H_\zeta H_\eta H_\zeta cos^2\theta cos^2\xi + V_\zeta V_\eta V_\zeta V_\eta sin^2\theta sin^2\xi - 2H_\eta H_\zeta V_\zeta V_\eta sin\theta sin\xi cos\theta cos\xi\right),$$

$$= \frac{I_0^2}{16}(cos\theta cos\xi - sin\theta sin\xi)^2,$$

$$= \frac{I_0^2}{16}\cos^2(\theta + \xi). \qquad (15)$$

Equation (15) witnesses the typical nonlocal quantum feature, satisfying the Bell inequality violation [29]. Such macroscopic nonlocal correlation has already been discussed recently [34]. Figure 5(b) is the corresponding numerical calculations of Eq. (15), where $R_{12}(0)$ is normalized. As shown in the bottom panel of Fig. 5(b), the ζ-dependent fringe shift is the definite proof of the nonlocal quantum feature [27,29]. On the contrary, Fig. 5(c) is for the classical intensity product for all product-basis terms for Eqs. (7) and (8), resulting in the typical classical feature of local realism. In Figs. 5(b) and (c) $\varphi = 0$ is set.

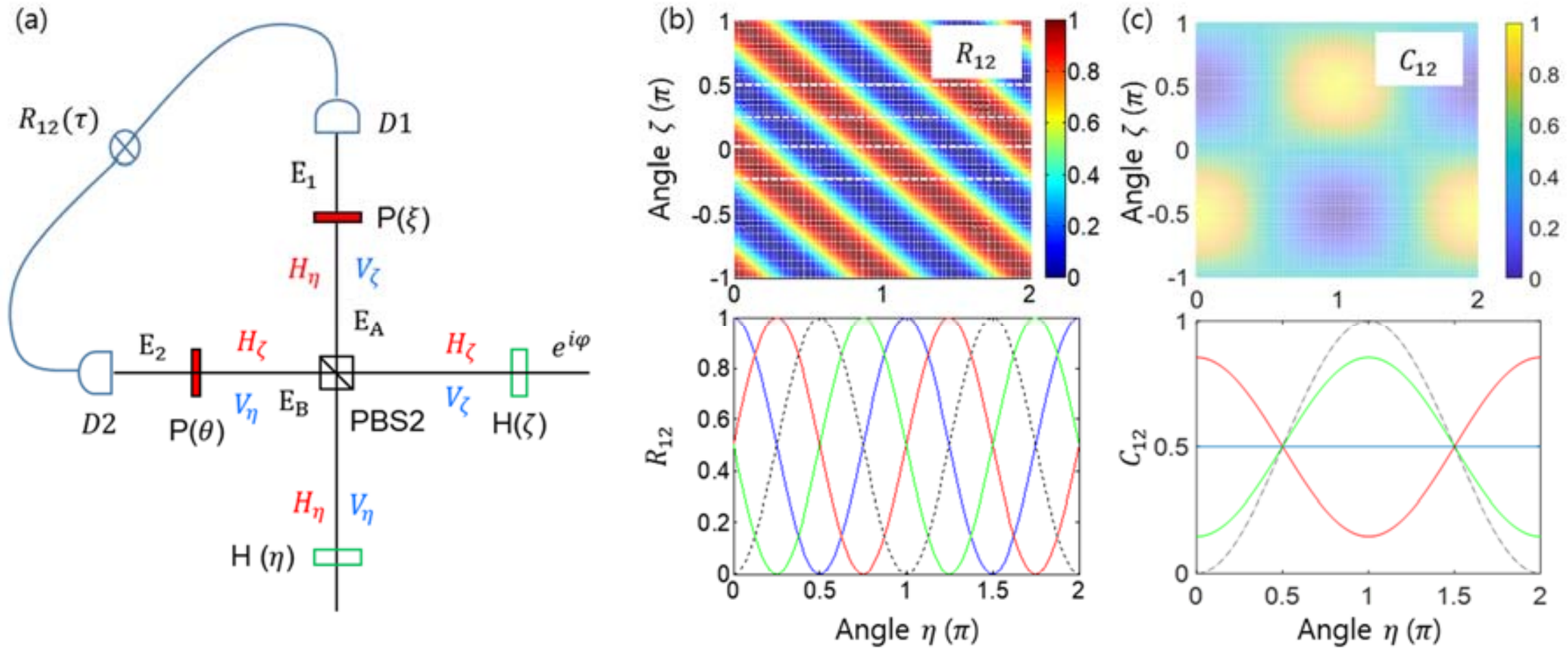

**Fig. 5.** Numerical calculations of intensity correlation. (a) Schematic of measurement. (b) Nonlocal correlation. (c) Classical intensity product. (blue) $\zeta = 0$, (red) $\zeta = -\pi/4$, (green) $\zeta = \pi/4$, (dotted) $\zeta = \pi/2$.

**Conclusion**

A macroscopic quantum eraser was experimentally demonstrated for the fundamental physics of the quantum mystery in the wave-particle duality using a cw laser. The preset photon characteristic of the particle nature in a noninterfering MZI was retrospectively switched to the wave nature via a measurement choice of the MZI output fields. Due to the quantum mechanical equality between quantum and coherence approaches for the first-order intensity correlation [10], the observed macroscopic quantum eraser has no distinction from the conventional single-photon-based quantum eraser [32]. Due to the delayed choice beyond the light cone, however, the causality issue was also raised as in the conventional microscopic regime. Although the same quantum mystery was involved in the macroscopic one, coherence has no direct relation to the light cone because of the relative length or time between two objects. Compared to the single photon-based quantum eraser, the observed macroscopic quantum eraser was deterministically understood as a selective choice of polarization bases via projection measurement. Moreover, the observed quantum eraser was discussed for *ad-hoc* quantum superposition between orthogonal polarization bases. Furthermore, the nonlocal quantum feature was also discussed for the intensity product between quantum erasers assuming selective product bases, where the quantum eraser is a typical method for nonlocal correlation measurements. Based on these understandings, the macroscopic quantum eraser can be applied to coherently controlled quantum technologies such as quantum spectrometers using macroscopic superresolution and various nonlocal quantum features by selectively choosing a particular set of product bases.

**Reference**

1. Dirac, P. A. M. The principles of Quantum mechanics. 4th ed. (Oxford University Press, London), Ch. 1, p. 9 (1958).
2. Feynman R P, Leighton R, and Sands M 1965 The Feynman Lectures on Physics, Vol. III (Addison Wesley, Reading, MA).
3. Wheeler, J. A. in *Mathematical Foundations of Quantum Theory*, Marlow, A. R. Ed. (Academic Press 1978), pp. 9-48.
4. Scully, M. O. and Drühl, K. Quantum eraser: A proposed photon correlation experiment concerning observation and “delayed choice” in quantum mechanics. *Phys. Rev.* A **25**, 2208-2213 (1982).
5. Scully, M. O., Englert, B.-G. and Walther, H. Quantum optical tests of complementarity. *Nature* **351**, 111-116 (1991).
6. Ou, Z. Y., Wang, L. J., Zou, X. Y. and Mandel, L. Evidence for phase memory in two-photon down-conversion through entanglement with the vacuum. *Phys. Rev.* A **41**, 566-568 (1990).
7. Herzog, T. J., Kwiat, P. G., Weinfurter, H. and Zeilinger, A. Complementarity and the quantum eraser. *Phys. Rev. Lett.* **75**, 3034-3037 (1995).
8. Kim, Y.-H., Yu, R., Kulik, S. P. and Shih, Y. Delayed “choice” quantum eraser. *Phys. Rev. Lett.* **84**, 1-5 (2000).
9. Sun, F. W. et al. Experimental demonstration of phase measurement precision beating standard quantum limit by projection measurement. EPL **82**, 24001 (2008).
10. Stöhr, J. Overcoming the diffraction limit by multi-photon interference: Tutorial. Adv. Opt. Photon. **11**, 215–313 (2019).
11. Rueckner, W. and Peidle, J. Young’s double-slit experiment with single photons and quantum eraser. *Am. J. Phys*. **81**, 951-958 (2013).
12. Kim, S. & Ham, B. S. Revisiting self-interference in Young's double-slit experiments. Sci. Rep. **13**, 977 (2023).
13. Henry, M. Fresnel-Arago laws for interference in polarized light: A demonstration experiment. *Am. J. Phys.* **49**, 690-691 (1981).
14. Grangier, P., Roger, G. and Aspect, A. Experimental evidence for a photon anticorrelation effect on a beam splitter: A new light on single-photon interferences. *Europhys. Lett.* **1**, 173-179 (1986).
15. Peng, T., Chen, H., Shih, Y. and Scully, M. O. Delayed-choice quantum eraser with thermal light. *Phys. Rev. Lett.* **112**, 180401 (2014).
16. Scully, M. O.; Englert, B.-G.; Walther, H. “Quantum optical tests of complementarity,” *Nature 351*, 111-116 (1991).
17. Aharonov, Y.; Zubairy, M. S. Time and the quantum: erasing the past and impacting the future. *Science **307***, 875-879 (2005).
18. Dimitrova, T. L. & Weis, A. Single photon quantum erasing: a demonstration expe4riment. Eur. J. Phys. **31**, 625 (2010).
19. Jacques, V., Wu, E., Grosshans, F., Treussart, F., Grangier, P., Aspect, A. and Roch, J.-F. Experimental realization of Wheeler’s delayed-choice gedanken experiments. *Science* **315**, 966-968 (2007).

20. Yu, S., Sun, Y.-N., Liu, W., Liu, Z.-D., Ke, Z.-J., Wang, Y.-T., Tang, J.-S., Li, C.-F. and Guo, G.-C. Realization of a causal-modeled delayed-choice experiment using single photons, *Phys. Rev.* A **100**, 012115 (2019).
21. Ma, X. S., Zotter, S., Kofler, J., Ursin, R., Jennewein, T., Brukner, C. and Zeilinger, A. Experimental delayed-choice entanglement swapping. *Nature Phys*. **8**, 479-484 (2012).
22. Dieguez, P. R. *et al.* Experimental assessment of physical realism on a quantum-controlled device. Communi. Phys. **5**, 82 (2022).
23. Ham, B. S. Coherently excited superresolution using intensity product of phase-controlled quantum erasers via polarization-basis projection measurements. Sci. Rep. **14**, 11521 (2024).
24. Kim, S. & Ham, B. S. Observations of super-resolution using phase-controlled coherent photons in a delayed-choice quantum eraser scheme. arXiv:2312.03343 (2023).
25. Giovannetti, V., Lloyd, S. & Lloyd, L. Quantum-enhanced measurement: beating the standard quantum limit. Science **306**, 1330-1336 (2004).
26. Nagata, T., Okamoto, R., O'Brian, J. L., Sasaki, K. & Takeuchi, S. Beating the standard quantum limit with four-entangled photons. Science **316**, 726–729 (2007).
27. Kim, T, Fiorentino, M, & Wong, F. N. C. Phase-stable source of polarization-entangled photons using a polarization Sagnac interferometer. *Phys. Rev. A* **73**, 012316 (2006).
28. Clauser, J. F., Horne, M. A., Shimony, S., & Holt, R. A. Proposed experiment to test local hidden-variable theories. Phys. Rev. Lett. **23**, 880-884 (1969).
29. Weihs, G., Jennewein, T., Simon, C., Weinfurter, H. & Zeilinger, A. Violation of Bell's inequality under strict Einstein locality conditions. Phys. Rev. Lett. **81**, 5039-5043 (1998).
30. Degiorgio, V. Phase shift between the transmitted and the reflected optical fields of a semirefecting lossless mirror is π/2. Am. J. Phys. **48**, 81–82 (1980).
31. Hardy, L. Source of photons with correlated polarizations and correlated directions. Phys. Lett. A **161**, 326-328 (1992).
32. Kim, S. and Ham, B. S. Observations of the delayed-choice quantum eraser using coherent photons. Sci. Rep. **13**, 9758 (2023).
33. Pedrotti, F. L., Pedrotti, L. M. & Pedrotti, L. S. *Introduction to Optics* 3rd edn. (Pearson Education Inc, 2004), Ch. 14.
34. Ham, B. S. Macroscopic quantum correlation using coherence manipulations of polarization-path correlations of a continuous-wave laser. arXiv:2308.04078 (2023).

**Acknowledgments**

This work was supported by the ICT R&D program of MSIT/IITP (2021-0-01810), development of elemental technologies for ultrasecure quantum internet. BSH also acknowledges that this work was also supported by the GIST Research Project in 2022.

**Conflict of Interests**

The author has no conflicts to disclose.

**Author contributions**

B.S.H. solely wrote the manuscript.